# EVALUATION OF THE COSMOLOGICAL CONSTANT IN INFLATION WITH A MASSIVE NON-MINIMAL SCALAR FIELD

JUNG-JENG HUANG
*Department of Mechanical Engineering, Physics Division*
*Ming Chi University of Technology, Taishan, New Taipei City 24301, Taiwan*
*huangjj@mail.mcut.edu.tw*

In Schrödinger picture we study the possible effects of trans-Planckian physics on the quantum evolution of massive non-minimally coupled scalar field in de Sitter space. For the nonlinear Corley-Jacobson type dispersion relations with quartic or sextic correction, we obtain the time evolution of the vacuum state wave functional during slow-roll inflation, and calculate explicitly the corresponding expectation value of vacuum energy density. We find that the vacuum energy density is finite. For the usual dispersion parameter choice, the vacuum energy density for quartic correction to the dispersion relation is larger than for sextic correction, while for some other parameter choices, the vacuum energy density for quartic correction is smaller than for sextic correction. We also use the backreaction to constrain the magnitude of parameters in nonlinear dispersion relation, and show how the cosmological constant depends on the parameters and the energy scale during the inflation at the grand unification phase transition.

## 1. Introduction

In the standard inflationary scenario, usual realization of inflation is associated with a slow rolling inflaton minimally coupled to gravity [1]. Nevertheless, it is well known that the extension to the non-minimal coupling with the Ricci scalar curvature can soften the problem related to the small value of the self-coupling in the quartic potential of chaotic inflation [2]. Further, non-minimal coupling terms also can lead to corrections on power spectrum of primordial perturbations [3], a tiny tensor-to-scalar ratio [4, 5] and non-Gaussianities [6]. A broad class of models of chaotic inflation in supergravity with an arbitrary inflaton potential was also proposed. In these models the inflaton field is non-minimally coupled to gravity [7, 8]. Recently, the viability of simple non-minimally coupled inflationary models is assessed through observational constraints on the magnitude of the non-minimal coupling from the BICEP2 experiment [9].

Moreover, the standard inflationary scenario has two possible extensions. The first extension is associated with the ambiguity of initial quantum vacuum state, and the choice of initial vacuum state affects the predictions of inflation [10, 11]. The second extension concerns with the trans-Planckian problem [12, 13] of whether the predictions of standard cosmology are insensitive to the effects of trans-Planckian physics. In fact, nonlinear dispersion relations such as the Corley-Jacobson (CJ) type were used to mimic the trans-Planckian effects on cosmological perturbations [12-14]. These CJ type dispersion relations can be obtained naturally from quantum gravity models such as Horava gravity [15, 16]. Recently, in several approaches to quantum gravity, the phenomenon of running spectral dimension of spacetime from the standard value of 4 in the infrared to a smaller value in the ultraviolet is associated with modified dispersion relations, which also include the CJ type dispersion relations [17, 18].

In the previous work [19-23] we used the lattice Schrödinger picture to study the free scalar field theory in de Sitter space, derived the wave functionals for the Bunch-Davies (BD) vacuum state and its excited states, and found the trans-Planckian effects on the quantum evolution of massless minimally coupled scalar field for the CJ type dispersion relations with sextic correction. In this paper we extend the study to the case of massive non-minimally coupled scalar field.

The paper is organized as follows. In Section 2, the theory of a generically coupled scalar field in de Sitter space is briefly reviewed in the lattice Schrödinger picture. In Section 3, we consider the massive non-minimally coupled scalar field during slow-roll inflation, and use the CJ type dispersion relations with quartic or sextic correction to obtain the time evolution of the vacuum state wave functional. In Section 4, using the results of Section 3, we calculate the finite vacuum energy density, use the backreaction to constraint the parameters in nonlinear dispersion, and evaluate the cosmological constant. Finally, conclusions and discussion are presented in Section 5. Throughout this paper we will set $\hbar$ =c=1.

## 2. De Sitter Scalar Field Theory in Schrödinger Picture

In this section, we begin by briefly reviewing the theory of a generically coupled scalar field in de Sitter space in the lattice Schrödinger picture (for the details of some derivations in this section see [23]). The Lagrangian density for the scalar field we consider is

$$L = |g|^{\frac{1}{2}}\left\{\frac{1}{2}\left[g^{\mu\nu}(x)\phi(x)_{,\mu}\phi(x)_{,\nu}\right] - \xi R\phi^2/2 - V(\phi)\right\},$$

$$V(\phi) = m^2\phi^2/2, \tag{1}$$

where $\phi$ is a real scalar field, $V(\phi)$ is the potential , $m$ is the mass of the scalar quanta, $R$ is the Ricci scalar curvature, $\xi$ is the coupling parameter, and $g = \det g_{\mu\nu}$, $\mu,\nu =$ 0,1,…,d. For a spatially flat (1+d)-dimensional Robertson-Walker space-time with scale factor $a(t)$, we have

$$ds^2 = dt^2 - a^2(t)d^2x^i,\ i = 1,2,...,d,$$
$$L = a^d\left\{\frac{1}{2}\left[(\partial_0\phi)^2 - a^{-2}(\partial_i\phi)^2\right] - \xi R\phi^2/2 - m^2\phi^2/2\right\}. \tag{2}$$

In the (1+d)-dimensional de Sitter space we have $a(t) = \exp(ht)$, where $h \equiv \dot{a}/a$ is the Hubble parameter which is a constant.

For d=1, in the lattice Schrödinger picture, we obtain from (2) the time-dependent functional Schrödinger equation in momentum space [23]

$$H\psi = i\frac{\partial}{\partial t}\psi, \tag{3}$$

where

$$H = 2\sum_{l=1}^{N/2}\sum_{r=1}^{2} H_{rl}, \tag{4}$$

$$H_{rl} = \frac{1}{2}p_{rl}^{\ 2} + \frac{1}{2}hp_{rl}\phi_{rl} + \frac{1}{2}a^{-2}\omega_l^{\ 2}\phi_{rl}^{\ 2} + \frac{1}{2}\left(m^2 + \xi R\right)\phi_{rl}^{\ 2}, \tag{5}$$

$$\psi[\phi_{rl}, t] = \prod_{l=1}^{N/2}\prod_{r=1}^{2}\psi_{rl}(\phi_{rl}, t) \equiv \prod_{rl}\psi_{rl}(\phi_{rl}, t), \tag{6}$$

Here $\omega_l \equiv (2/\varepsilon)\sin(l\pi/N), \varepsilon = W/N$, *i.e.*, $W$ is the overall comoving spatial size of lattice, $\phi_l = \phi_{1l} + i\phi_{2l}$, $p_l = p_{1l} + ip_{2l}$, $p_l$ is the conjugate momentum for $\phi_l$, and the subscripts 1 and 2 denote the real and imaginary parts respectively.

For each real mode $\phi_{rl}$, we have

$$H_{rl}\psi_{rl} = i\frac{\partial}{\partial t}\psi_{rl},\ r=1,2 \tag{7}$$

$$-\frac{1}{2}\frac{\partial^2\psi_{rl}}{\partial\phi_{rl}^{\ 2}} + \frac{1}{2}\left[a^{-2}\omega_l^{\ 2} + \left(m^2 + \xi R\right) - \frac{1}{4}h^2\right]\phi_{rl}^{\ 2}\psi_{rl} = i\frac{\partial\psi_{rl}}{\partial t}. \tag{8}$$

Note that (8) arises from the field quantization of the Hamiltonian (5) through the functional Schrödinger representation $\bar{\phi}_{rl} \to \phi_{rl}$, $\bar{p}_{rl} \to -i\partial/\partial\phi_{rl}$, where operators $\bar{\phi}_{rl}$ and $\bar{p}_{rl}$ satisfy the equal time commutation relations $[\bar{\phi}_{rl}, \bar{p}_{rl}] = i$, and setting $\bar{P}_{rl} = \bar{p}_{rl} + \frac{1}{2}h\bar{\phi}_{rl} = -i\partial/\partial\phi_{rl}$ so that $\left[\bar{\phi}_{rl}, \bar{P}_{rl}\right] = \left[\bar{\phi}_{rl}, \bar{p}_{rl}\right] = i$. Thus (8) governs the time evolution of the state wave functional $\left|\psi_{rl}\right\rangle$ of the Hamiltonian operator $H_{rl}$ in the $\{\left|\phi_{rl}\right\rangle\}$ representation. In terms of the conformal time $\tau$ defined by

$$d\tau = dt/a,\ \ \tau = -h^{-1}\exp(-ht) = -h^{-1}a^{-1},\ \ -\infty < \tau < 0, \tag{9}$$

the normalized wave functionals of vacuum and its excited states are

$$\psi_{rl(n_{rl})}(\phi_{rl},\tau)=R_{(n_{rl})}(\phi_{rl},\tau)\exp\left(i\Theta_{(n_{rl})}(\phi_{rl},\tau)\right),\ \ n_{rl}=0,1,2,... \quad (10)$$

with the amplitude $R_{(n_{rl})}(\phi_{rl},\tau)$ and phase $\Theta_{(n_{rl})}(\phi_{rl},\tau)$

$$R_{(n_{rl})}(\phi_{rl},\tau)=\left[\frac{\sqrt{2h/\pi}}{\sqrt{\pi}2^{n_{rl}}n_{rl}!\left|H_{\nu}^{(1)}\right|}\right]^{1/2}H_{(n_{rl})}(\eta_{rl})\exp(-\frac{1}{2}\eta_{rl}^{2}), \quad (11)$$

$$\Theta_{(n_{rl})}(\phi_{rl},\tau)=-\frac{h\omega_l|\tau|}{2}\frac{\left(\left|H_{\nu}^{(1)}\right|\right)'}{\left|H_{\nu}^{(1)}\right|}\phi_{rl}^{2}-(\frac{1}{2}+n_{rl})\int\frac{\frac{2}{\pi|\tau|}}{\left|H_{\nu}^{(1)}\right|^{2}}d\tau. \quad (12)$$

Here $\eta_{rl}$ is defined by $\eta_{rl}\equiv\left(\sqrt{2h/\pi}/\left|H_{\nu}^{(1)}\right|\right)\phi_{rl}$, $H_{(n_{rl})}(\eta_{rl})$ is the $n_{rl}$ th-order Hermite polynomial, $H_{\nu}^{(1)}(\omega_l|\tau|)$ is the Hankel function of the first kind of order $\nu$, $\nu^2=1/4-\left(m^2+\xi R\right)/h^2$, and the prime in (12) denotes the derivative with respect to $\omega_l|\tau|$. The complete wave functionals can be written as $\psi_{[n]}[\phi_{rl},t]=\prod_{rl}\psi_{(n_{rl})}(\phi_{rl},t)$, where $[n]\equiv(n_i,n_j,\cdots)$ means that mode $i$ is in the $n_i$ excited state, mode $j$ is in the $n_j$ excited state, *etc.* For $n_{rl}=0$, the ground state wave functional corresponds to the BD vacuum.

For d=3, we have $\nu^2=9/4-\left(m^2+\xi R\right)/h^2$, $R=12h^2$ and the mode index $l$ in $\omega_l$ carries labels $(l_i, i=1,2,3)$ which will be suppressed below. Furthermore, from equations (3)-(8) we get in the continuum limit ( $\omega_l\to k$ )

$$i\frac{\partial\psi}{\partial t}=\sum_{rk}\{-\frac{1}{2}\frac{\partial^2}{\partial\phi_{rk}^{2}}+\frac{1}{2}\left[a^{-2}k^2+\left(m^2+\xi R\right)-\frac{9}{4}h^2\right]\phi_{rk}^{2}\}\psi. \quad (13)$$

**3. Trans-Planckian Effects on Vacuum Wave Functional**

For the inflationary potential $V(\phi)=m^2\phi^2/2$, the bounds on $\xi$ derived from the joint data analysis of Planck+WP+BAO+high-*l* for the number of e-foldings $N=60$ are $-4.2\times10^{-3}<\xi<-1.1\times10^{-3}$ (68%CL), $-5.1\times10^{-3}<\xi\le0$ (95%CL) [24].

For the mass in the tree-level potential, we have $m = 1.46\times10^{13}$ GeV [25]. Moreover, the recent BICEP2 experiment suggests that $h \cong 1.13\times10^{14}$ GeV [26-28]. Therefore, from $\nu^2 = 9/4 - \left(m^2 + \xi R\right)/h^2$, we have $m^2 + \xi R << 9h^2/4$ and $\nu \cong 3/2$, which will be used below. To study further the effects of trans-Planckian physics, we use the CJ type dispersion relations

$$\omega^2(k/a) = k^2\left[1 + b_s(\frac{k}{aM})^{2s}\right], \tag{14}$$

where $M$ is a cutoff scale, $s$ is an integer, and $b_s$ is an arbitrary coefficient [12-14].

*3.1. CJ Type Dispersion Relations with Quartic Correction*

First, we use the CJ type dispersion relations (14) with $s = 1$ and $b_1 > 0$ to obtain the time evolution of the vacuum state wave functional. Recall that these CJ type dispersion relations can be obtained from theories based on quantum gravity models [15-18].

Using $z = k|\tau| = k/ah$ which is the ratio of physical wave number $k_{phys} \equiv k/a$ to the inverse of Hubble radius, (13) becomes

$$i\frac{\partial\psi}{\partial t} = \sum_{rk}\{-\frac{1}{2}\frac{\partial^2}{\partial\phi_{rk}{}^2} + \frac{1}{2}\left[z^2\left(1+\sigma^2 z^2\right)h^2 - \frac{9}{4}h^2\right]\phi_{rk}{}^2\}\psi, \tag{15}$$

where $\sigma^2 \equiv b_1(h/M)^2$, and the ground state solution of (15) becomes

$$\psi_{(0)} = \prod_{rk} A_{k(0)}(\tau)\exp(-\frac{1}{2}B_k(\tau)a^{-1}\phi_{rk}{}^2), \tag{16}$$

where $A_{k(0)}(\tau)$ and $B_k(\tau)$ satisfy

$$A_{k(0)}(\tau) = \exp\left[-i\frac{1}{2}\int B_k(\tau)d\tau + const\right], \tag{17}$$

$$B_k{}^2(\tau) - i\left[\frac{dB_k(\tau)}{d\tau} + \frac{B_k(\tau)}{\tau}\right] - \left[k^2(1+\sigma^2 z^2) - \frac{9}{4\tau^2}\right] = 0. \tag{18}$$

In region I where $k_{phys} \equiv k/a > M$, *i.e.*, $z > M/h$, the dispersion relations can be approximated by $\omega^2(k/a) \approx k^2\sigma^2 z^2$, and the corresponding wave functional for the initial BD vacuum state is [23, 29]

$$\psi_{(0)}{}^{\mathrm{I}} = \prod_{rk} A_{k(0)}{}^{\mathrm{I}}(\tau)\exp(-\frac{1}{2}B_k{}^{\mathrm{I}}(\tau)a^{-1}\phi_{rk}{}^{\mathrm{I}2}),$$

$$A_{k(0)}{}^{\mathrm{I}}(\tau) = \exp\left[-i\frac{1}{2}\int B_k{}^{\mathrm{I}}(\tau)d\tau + const\right], \tag{19}$$

$$B_k{}^{\mathrm{I}}(\tau) = \frac{\dfrac{4}{\pi|\tau|}}{\left|H_{3/4}{}^{(1)}\right|^2} - i\frac{k}{2}\frac{\left(\left|H_{3/4}{}^{(1)}\right|^2\right)'}{\left|H_{3/4}{}^{(1)}\right|^2}\sigma z, \tag{20}$$

where the prime in (20) denotes the derivative with respect to $\sigma z^2/2$.

On the other hand, in region II where $k_{phys} \equiv k/a < M$, *i.e.*, $z < M/h$, linear relations recover $\omega^2 \cong k^2$, and the corresponding wave functional for the non-BD vacuum state is [23, 29]

$$\psi_{(0)}{}^{\mathrm{II}} = \prod_{rk} A_{k(0)}{}^{\mathrm{II}}(\tau)\exp(-\frac{1}{2}B_k{}^{\mathrm{II}}(\tau)a^{-1}\phi_{rk}{}^{\mathrm{II}2}),$$

$$A_{k(0)}{}^{\mathrm{II}}(\tau) = \exp\left[-i\frac{1}{2}\int B_k{}^{\mathrm{II}}(\tau)d\tau + const\right], \tag{21}$$

$$B_k{}^{\mathrm{II}}(\tau) = \frac{\dfrac{2}{\pi|\tau|}}{\left(\left|C_1{}^{\mathrm{II}}\right|^2 + \left|C_2{}^{\mathrm{II}}\right|^2\right)\left|H_{3/2}{}^{(1)}\right|^2 + 2\,\mathrm{Re}\left[C_1{}^{\mathrm{II}}C_2{}^{\mathrm{II}*}\left(H_{3/2}{}^{(1)}\right)^2\right]}$$

$$-i\frac{k}{2}\frac{\left\{\left(\left|C_1{}^{\mathrm{II}}\right|^2 + \left|C_2{}^{\mathrm{II}}\right|^2\right)\left|H_{3/2}{}^{(1)}\right|^2 + 2\,\mathrm{Re}\left[C_1{}^{\mathrm{II}}C_2{}^{\mathrm{II}*}\left(H_{3/2}{}^{(1)}\right)^2\right]\right\}'}{\left(\left|C_1{}^{\mathrm{II}}\right|^2 + \left|C_2{}^{\mathrm{II}}\right|^2\right)\left|H_{3/2}{}^{(1)}\right|^2 + 2\,\mathrm{Re}\left[C_1{}^{\mathrm{II}}C_2{}^{\mathrm{II}*}\left(H_{3/2}{}^{(1)}\right)^2\right]}, \tag{22}$$

where the prime in (22) denotes the derivative with respect to $z$, and $C_1{}^{\mathrm{II}}$ and $C_2{}^{\mathrm{II}}$ satisfy $\left|C_1{}^{\mathrm{II}}\right|^2 - \left|C_2{}^{\mathrm{II}}\right|^2 = 1$. Let $\tau_c$ be the time when the modified dispersion relations take the standard linear form. Then $\sigma^2 z_c{}^2 = 1$ where $z_c = k|\tau_c| = M/b_1^{1/2}h$ >>1 for $b_1 \sim 1$. The constants $C_1^{\text{II}}$ and $C_2^{\text{II}}$ can be obtained by the following matching conditions at $\tau_c$ for the two wave functionals (19) and (21)

$$\psi_{(0)}{}^{\text{I}}\Big|_{z_C} = \psi_{(0)}{}^{\text{II}}\Big|_{z_C}\,, \tag{23}$$

$$\frac{d\psi_{(0)}{}^{\text{I}}}{dz}\Big|_{z_C} = \frac{d\psi_{(0)}{}^{\text{II}}}{dz}\Big|_{z_C}\,, \tag{24}$$

which can also be rewritten respectively as

$$\operatorname{Re}\left(B_k{}^{\text{I}}\right)\Big|_{z_C} = \operatorname{Re}\left(B_k{}^{\text{II}}\right)\Big|_{z_C}\,, \tag{25}$$

$$\frac{d\operatorname{Re}\left(B_k{}^{\text{I}}\right)}{dz}\Big|_{z_C} = \frac{d\operatorname{Re}\left(B_k{}^{\text{II}}\right)}{dz}\Big|_{z_C}\,, \tag{26}$$

by requiring $B_k{}^{\text{I}} = B_k{}^{\text{II}}$, $\phi_{rk}{}^{\text{I}} = \phi_{rk}{}^{\text{II}}$, and $A_{k(0)}{}^{\text{I}} = A_{k(0)}{}^{\text{II}}$ when $z = z_c$.

Using $\left|H_{3/4}{}^{(1)}\left(\sigma z^2/2\right)\right|^2 = \left(4/\pi\sigma z^2\right)\left(1+5/8\sigma^2 z^4+\ldots\right) \approx 4/\pi\sigma z^2$ with $\sigma = z_c{}^{-1}$ and $z_c >> 1$, we have from (20), (22), and (25)

$$1 = \left|C_1^{\text{II}}\right|^2 + \left|C_2^{\text{II}}\right|^2 + 2\left|C_1^{\text{II}}\right|\left|C_2^{\text{II}}\right|\cos(2z_c - \theta)\,, \tag{27}$$

where we choose $C_1^{\text{II}} = \left|C_1^{\text{II}}\right|$ and $C_2^{\text{II}} = \left|C_2^{\text{II}}\right|\exp(i\theta)$, and $\theta$ is a relative phase parameter. Then from (27) and $\left|C_1^{\text{II}}\right|^2 - \left|C_2^{\text{II}}\right|^2 = 1$ we have

$$\left|C_1^{\text{II}}\right| = \csc\left(2z_c - \theta\right)\ ,\ \left|C_2^{\text{II}}\right| = -\cot(2z_c - \theta)\,, \tag{28}$$

where $\sin(2z_c - \theta) > 0$, $\cos(2z_c - \theta) < 0$. Substituting (20) and (22) into (26) and keeping terms up to order $1/z_c$ on the right-hand side of (26) , we obtain

$$\frac{1}{z_c} = \left|C_1^{\text{II}}\right|\left|C_2^{\text{II}}\right|\cos(2z_c - \theta)\frac{8}{z_c} + 4\left|C_1^{\text{II}}\right|\left|C_2^{\text{II}}\right|\sin\left(2z_c - \theta\right). \tag{29}$$

Using (28) in (29) gives

$$\cot(2z_c-\theta)=-\frac{1}{4z_c} \quad \text{or} \quad \cot(2z_c-\theta)=-\frac{z_c}{2}+\frac{1}{4z_c}. \tag{30}$$

Here we choose $\cot(2z_c-\theta)=-\frac{1}{4z_c}$, so that $\left|C_2^{\,\mathrm{II}}\right|$ is small for $z_c$ >>1 to avoid an unacceptably large backreaction on the background geometry. Then we have

$$\left|C_2^{\,\mathrm{II}}\right|\cong\frac{1}{4z_c},\ \left|C_1^{\,\mathrm{II}}\right|=\sqrt{1+\left|C_2^{\,\mathrm{II}}\right|^2}\cong 1+\frac{1}{32{z_c}^2}\cong 1, \tag{31}$$

or

$$\sin(2z_c-\theta)\cong 1,\ \cos(2z_c-\theta)\cong-\frac{1}{4z_c}. \tag{32}$$

*3.2. CJ Type Dispersion Relations with Sextic Correction*

In this subsection, we use the CJ type dispersion relations (14) with $s=2$ and $b_2>0$ to obtain the time evolution of the vacuum state wave functional. For this case, only (15), (18), and (20) are changed into

$$i\frac{\partial\psi}{\partial t}=\sum_{rk}\{-\frac{1}{2}\frac{\partial^2}{\partial{\phi_{rk}}^2}+\frac{1}{2}\left[z^2\left(1+\sigma^2z^4\right)h^2-\frac{9}{4}h^2\right]{\phi_{rk}}^2\}\psi, \tag{33}$$

$$B_k^{\,2}(\tau)-i\left[\frac{dB_k(\tau)}{d\tau}+\frac{B_k(\tau)}{\tau}\right]-\left[k^2(1+\sigma^2z^4)-\frac{9}{4\tau^2}\right]=0, \tag{34}$$

$$B_k^{\,\mathrm{I}}(\tau)=\frac{\dfrac{6}{\pi|\tau|}}{\left|H_{1/2}^{(1)}\right|^2}-i\frac{k}{2}\frac{\left(\left|H_{1/2}^{(1)}\right|^2\right)'}{\left|H_{1/2}^{(1)}\right|^2}\sigma z^2, \tag{35}$$

where $\sigma^2\equiv b_2(h/M)^4$, and the prime in (35) denotes the derivative with respect to $\sigma z^3/3$. Using $\left|H_{1/2}^{(1)}\left(\sigma z^3/3\right)\right|^2=\left(6/\pi\sigma z^3\right)$ with $\sigma=\bar{z}_c^{\,-2}$ and $\bar{z}_c=k|\bar{\tau}_c|=M/b_2^{1/4}h$ >>1 for $b_2\sim 1$, we obtain from (35), (22), (25) and $\left|C_1^{\,\mathrm{II}}\right|^2-\left|C_2^{\,\mathrm{II}}\right|^2=1$

$$1=\left|C_1^{\,\mathrm{II}}\right|^2+\left|C_2^{\,\mathrm{II}}\right|^2+2\left|C_1^{\,\mathrm{II}}\right|\left|C_2^{\,\mathrm{II}}\right|\cos(2\bar{z}_c-\theta), \tag{36}$$

$$\left|C_1^{\mathrm{II}}\right| = \csc\left(2\bar{z}_c - \theta\right) \ , \ \left|C_2^{\mathrm{II}}\right| = -\cot(2\bar{z}_c - \theta), \tag{37}$$

where $\sin(2\bar{z}_c - \theta) > 0$, $\cos(2\bar{z}_c - \theta) < 0$. Substituting (35) and (22) into (26) and keeping terms up to order $1/\bar{z}_c$ on the right-hand side of (26), we find

$$\frac{2}{\bar{z}_c} = \left|C_1^{\mathrm{II}}\right|\left|C_2^{\mathrm{II}}\right|\cos(2\bar{z}_c - \theta)\frac{8}{\bar{z}_c} + 4\left|C_1^{\mathrm{II}}\right|\left|C_2^{\mathrm{II}}\right|\sin\left(2\bar{z}_c - \theta\right). \tag{38}$$

Using (37) in (38) gives

$$\cot(2\bar{z}_c - \theta) = -\frac{1}{2\bar{z}_c} \quad \text{or} \quad \cot(2\bar{z}_c - \theta) = -\frac{\bar{z}_c}{2} + \frac{1}{2\bar{z}_c}. \tag{39}$$

Here we choose $\cot(2\bar{z}_c - \theta) = -\frac{1}{2\bar{z}_c}$, so that $\left|C_2^{\mathrm{II}}\right|$ is small for $\bar{z}_c$ >>1 to avoid an unacceptably large backreaction on the background geometry. Then we have

$$\left|C_2^{\mathrm{II}}\right| \cong \frac{1}{2\bar{z}_c}, \ \left|C_1^{\mathrm{II}}\right| = \sqrt{1 + \left|C_2^{\mathrm{II}}\right|^2} \cong 1 + \frac{1}{8\bar{z}_c^{\ 2}} \cong 1, \tag{40}$$

or

$$\sin(2\bar{z}_c - \theta) \cong 1, \ \cos(2\bar{z}_c - \theta) \cong -\frac{1}{2\bar{z}_c}. \tag{41}$$

## 4. Vacuum Energy, Backreaction and Cosmological Constant

Using the results of Section 3, we proceed to calculate the finite vacuum energy density and use the backreaction constraint to address the cosmological constant problem. Note that in the slow-roll approximation, the energy density of the scalar field is $\rho_\phi \cong V(\phi)$, where $V(\phi) = m^2\phi^2/2$. Therefore the relation between the expectation value of the vacuum energy density $\rho_\phi$ and the vacuum wave functional $\psi_{(0)}$ in (16) is

$$\begin{aligned}\left\langle \rho_\phi \right\rangle = \left\langle \psi_{(0)} \middle| \rho_\phi \middle| \psi_{(0)} \right\rangle &= \frac{m^2}{2}\varepsilon^{-3}\sum_{rk}\int_{-\infty}^{\infty} du_{rk}\left|\psi_{rk(0)}(u_{rk},\tau)\right|^2 u_{rk}^{\ 2} \\ &= \frac{m^2}{2}\frac{1}{8\pi^3}\int d^3k \frac{1}{2}\frac{1}{\mathrm{Re}(B_k(\tau)a^{-1})}a^{-3} \\ &= \frac{m^2}{2}\frac{1}{2\pi^2}\int k^2 \frac{1}{2}\frac{1}{\mathrm{Re}(B_k(\tau)a^{-1})}a^{-3}dk\,,\end{aligned} \tag{42}$$

where we use a field redefinition $u_{rk} \equiv a^{-3/2}\phi_{rk}$,

$$\left|\psi_{rk(0)}(u_{rk},\tau)\right|^2 = a^{3/2}\,\frac{\sqrt{\mathrm{Re}(B_k(\tau)a^{-1})}}{\sqrt{\pi}}\exp\left(-\mathrm{Re}(B_k(\tau)a^{-1})\frac{u_{rk}{}^2}{a^{-3}}\right), \qquad (43)$$

$\mathrm{Re}(B_k(\tau)a^{-1})$ denotes the real part of $B_k(\tau)a^{-1}$, and the factor $a^{3/2}$ in (43) appears through the normalization condition

$$\int_{-\infty}^{\infty} du_{rk}\left|\psi_{rk(0)}(u_{rk},\tau)\right|^2 = 1. \qquad (44)$$

For $s=1$ and $b_1>0$, in region I, we have $\left|H_{3/4}{}^{(1)}\left(\sigma z^2/2\right)\right|^2 \approx 4/\pi\sigma z^2$ with $\sigma = z_c{}^{-1}$. Then, using $a = 1/h|\tau| = k/hz$ and (20) in (42), we obtain

$$\left\langle \rho_\phi \right\rangle_{s=1}^{\mathrm{I}} = \frac{1}{8\pi^2}m^2h^2 z_c\int_{z_c}^{\alpha z_c} dz = \frac{1}{8\pi^2}m^2h^2 z_c{}^2(\alpha-1), \qquad (45)$$

where $z_c = M/b_1^{1/2}h$ and $\alpha z_c = M_{Pl}/h$ ($M_{Pl} = G^{-1/2} = 1.22\times10^{19}$ GeV is the Planck mass) are the boundaries of the interval of integration. On the other hand, in region II, (22) can be expressed as

$$B_k{}^{\mathrm{II}}(\tau) = \frac{\dfrac{2}{\pi|\tau|}}{\left|H_{3/2}{}^{(1)}\right|_{md}^2} - i\frac{k}{2}\frac{\left(\left|H_{3/2}{}^{(1)}\right|_{md}^2\right)'}{\left|H_{3/2}{}^{(1)}\right|_{md}^2}, \qquad (46)$$

where $\left|H_{3/2}{}^{(1)}\right|_{md}$ is defined as

$$\left|H_{3/2}{}^{(1)}\right|_{md} \equiv \left\{\left(\left|C_1{}^{\mathrm{II}}\right|^2 + \left|C_2{}^{\mathrm{II}}\right|^2\right)\left|H_{3/2}{}^{(1)}\right|^2 + 2\,\mathrm{Re}\left[C_1{}^{\mathrm{II}}C_2{}^{\mathrm{II}*}\left(H_{3/2}{}^{(1)}\right)^2\right]\right\}^{1/2}, \qquad (47)$$

with $\left|H_{3/2}{}^{(1)}(z)\right|^2 = \dfrac{2}{\pi z}\left(1+\dfrac{1}{z^2}\right)$. From (27), (31), and (32), we note that $\left|H_{3/2}{}^{(1)}\right|_{md}$ can be approximated by $\left|H_{3/2}{}^{(1)}\right|$ as $z$ decreases from $z = z_c >> 1$ to $z = 1$ (horizon exit). Then, using $a = 1/h|\tau| = k/hz$ and (22) in (42), we obtain

$$\left\langle \rho_\phi \right\rangle_{s=1}^{\mathrm{II}} = \frac{1}{8\pi^2} m^2 h^2 \int_1^{z_c} \frac{z^2+1}{z} dz = \frac{1}{8\pi^2} m^2 h^2 \left( \frac{1}{2} z_c{}^2 + \ln z_c - \frac{1}{2} \right). \qquad (48)$$

From (45) and (48) we have

$$\left\langle \rho_\phi \right\rangle_{s=1} = \left\langle \rho_\phi \right\rangle_{s=1}^{\mathrm{I}} + \left\langle \rho_\phi \right\rangle_{s=1}^{\mathrm{II}} = \frac{1}{8\pi^2} m^2 h^2 \left[ z_c{}^2 (\alpha - \frac{1}{2}) + \ln z_c - \frac{1}{2} \right], \qquad (49)$$

For $z_c >> 1$ and $\alpha = b_1^{1/2} (M_{Pl} / M) > 3/2$, (49) becomes

$$\left\langle \rho_\phi \right\rangle_{s=1} \cong \frac{1}{8\pi^2} m^2 h^2 z_c{}^2 \alpha . \qquad (50)$$

From (50) we see that there is no backreaction problem if the energy density due to the quantum fluctuations of the inflaton field is smaller than that due to the inflaton potential, *i.e.*,

$$\left\langle \rho_\phi \right\rangle_{s=1} < V(\phi) . \qquad (51)$$

In the slow-roll approximation, using $V(\phi) \cong 3 M_{Pl}{}^2 h^2 / 8\pi$ and (50) in (51) gives the constraint on the parameter $b_1$ as $b_1 > \frac{1}{9\pi^2} \left( \frac{M}{M_{Pl}} \right)^2 \left( \frac{m}{h} \right)^4$. For $M \sim 10^{16}$ GeV (the energy scale during inflation implied by the BICEP2 experiment [26, 27]), we have $b_1 > 2.1 \times 10^{-12}$.

For $s = 2$ and $b_2 > 0$, in region I, we have $\left| H_{1/2}{}^{(1)} \left( \sigma z^3 / 3 \right) \right|^2 = 6 / \pi \sigma z^3$ with $\sigma = \bar{z}_c{}^{-2}$. Then, using $a = 1 / h |\tau| = k / hz$ and (35) in (42), we obtain

$$\left\langle \rho_\phi \right\rangle_{s=2}^{\mathrm{I}} = \frac{1}{8\pi^2} m^2 h^2 \bar{z}_c{}^2 \int_{\bar{z}_c}^{\beta \bar{z}_c} \frac{1}{z} dz = \frac{1}{8\pi^2} m^2 h^2 \bar{z}_c{}^2 \ln \beta , \qquad (52)$$

where $\bar{z}_c = M / b_2^{1/4} h$ and $\beta \bar{z}_c = M_{Pl} / h$ are the boundaries of the interval of integration. On the other hand, in region II, (22) can be again expressed as (46) with $\left| H_{3/2}{}^{(1)} \right|_{md}$ defined by (47). From (36), (40), and (41), we also note that $\left| H_{3/2}{}^{(1)} \right|_{md}$ can be approximated by $\left|H_{3/2}{}^{(1)}\right|$ as $z$ decreases from $z = \bar{z}_c >> 1$ to $z = 1$. Then, using $a = 1/h|\tau| = k/hz$ and (22) in (42), we obtain

$$\left\langle \rho_\phi \right\rangle_{s=2}^{\mathrm{II}} = \frac{1}{8\pi^2} m^2 h^2 \int_1^{\bar{z}_c} \frac{z^2+1}{z} dz = \frac{1}{8\pi^2} m^2 h^2 \left( \frac{1}{2} \bar{z}_c{}^2 + \ln \bar{z}_c - \frac{1}{2} \right). \tag{53}$$

From (52) and (53) we have

$$\left\langle \rho_\phi \right\rangle_{s=2} = \left\langle \rho_\phi \right\rangle_{s=2}^{\mathrm{I}} + \left\langle \rho_\phi \right\rangle_{s=2}^{\mathrm{II}} = \frac{1}{8\pi^2} m^2 h^2 \left[ \bar{z}_c{}^2 (\ln \beta + \frac{1}{2}) + \ln \bar{z}_c - \frac{1}{2} \right]. \tag{54}$$

For $\bar{z}_c >> 1$ and $\bar{z}_c{}^2 (\ln \beta + 1/2) >> \ln \bar{z}_c$ which are satisfied if $b_2 > 6.3 \times 10^{-14}$, (54) becomes

$$\left\langle \rho_\phi \right\rangle_{s=2} \cong \frac{1}{8\pi^2} m^2 h^2 \bar{z}_c{}^2 (\ln \beta + \frac{1}{2}). \tag{55}$$

Moreover, we notice that there is no backreaction problem if

$$\left\langle \rho \right\rangle_{s=2} < V(\phi). \tag{56}$$

Using $V(\phi) \cong 3 M_{Pl}{}^2 h^2 / 8\pi$ and (55) in (56) gives the constraint on the parameter $b_2$ as $b_2{}^{1/2} > \frac{1}{3\pi} (\frac{Mm}{M_{Pl} h})^2 (\ln \beta + \frac{1}{2})$. For $M \sim 10^{16}$ GeV, we have $b_2 > 2.0 \times 10^{-7}$.

Comparing (50) with (54), we find that $\left\langle \rho_\phi \right\rangle_{s=2} < \left\langle \rho_\phi \right\rangle_{s=1}$ if the inequality $\bar{z}_c{}^2 (\ln \beta + 1/2) < z_c{}^2 \alpha$, or $\left[ (\ln \beta + 1/2)/(M_{Pl}/M) \right]^2 < b_2 / b_1$ is satisfied. For example, the usual parameter choice $b_1 \sim b_2 \sim 1$ satisfies the inequality. On the other hand, we have $\left\langle \rho_\phi \right\rangle_{s=2} > \left\langle \rho_\phi \right\rangle_{s=1}$ if the inequality $\bar{z}_c{}^2 (\ln \beta + 1/2) > z_c{}^2 \alpha$, or $\left[ (\ln \beta + 1/2)/(M_{Pl}/M) \right]^2 > b_2 / b_1$ is satisfied. For example, the parameter choice $b_1 \sim 1$ and $b_2 \sim 10^{-6}$ satisfies the inequality.

In the case that $\left\langle \rho_\phi \right\rangle_{s=1}$ is larger than $\left\langle \rho_\phi \right\rangle_{s=2}$, using (50) in the cosmological constant $\Lambda = 8\pi\rho_{vac} / M_{Pl}{}^2$ gives $\Lambda = \frac{1}{\pi b_1^{1/2}}\left(\frac{m^2 M}{M_{Pl}}\right)$ which is $5.6\times10^{22}\,\text{GeV}^2$ for $b_1 \sim 1$ and $1.8\times10^{28}\,\text{GeV}^2$ for $b_1 \sim 10^{-11}$. In the case that $\langle\rho_\phi\rangle_{s=2}$ is larger than $\langle\rho_\phi\rangle_{s=1}$, using (54) in the cosmological constant $\Lambda = 8\pi\rho_{vac} / M_{Pl}{}^2$ gives $\Lambda = \frac{1}{\pi b_2^{1/2}}\left(\frac{m^2 M^2}{M_{Pl}{}^2}\right)\left(\ln\beta + \frac{1}{2}\right)$ which is $3.5\times10^{20}\,\text{GeV}^2$ for $b_2 \sim 1$ and $1.9\times10^{23}\,\text{GeV}^2$ for $b_2 \sim 10^{-6}$.

**5. Conclusions and Discussion**

In the Schrödinger picture, we have considered the theory of a generically coupled free real scalar field in de Sitter space. To investigate the possible effects of trans-Planckian physics on the quantum evolution of the vacuum state of scalar field, we focus on the massive non-minimally coupled scalar field in slow-roll inflation, and consider the CJ type dispersion relations with quartic or sextic correction.

We obtain the time evolution of the vacuum state wave functional, and calculate the expectation value of the corresponding vacuum energy density. We find that the vacuum energy density is finite and has improved ultraviolet properties. For the usual dispersion parameter choice, the vacuum energy density for quartic correction to the dispersion relation is larger than for sextic correction. For some other parameter choices, the vacuum energy density for quartic correction is smaller than for sextic correction.

We also use the backreaction to constrain the magnitude of parameters in nonlinear dispersion relation, and show how the cosmological constant depends on the parameters and the energy scale during the inflation at the grand unification phase transition.

From (50) and (54) we see that the value of the cosmological constant can be reduced significantly through increasing the dispersion parameters in nonlinear dispersion relation and decreasing the cutoff energy scale associated with phase transition. However, the fact that the dispersion relation of a scalar field can not be modified on energy scales small than 1TeV makes the cosmological problem still unsolved.

**Acknowledgments**

The author thanks M.-J. Wang for stimulating discussions on the cosmological constant, and his colleagues at Ming Chi University of Technology for useful suggestions.